\documentclass[conference,10pt,twocolumn,twoside]{IEEEtran}
\IEEEoverridecommandlockouts
\usepackage{amsmath}
\usepackage{amsfonts}
\usepackage{cases}
\usepackage{setspace}
\usepackage{fancybox}
\usepackage{subfigure}
\usepackage{epsfig}
\usepackage{graphicx}
\usepackage{epstopdf}
\usepackage{float}
\usepackage{multirow}
\usepackage{color}
\usepackage{amsmath}
\usepackage{multirow}
\usepackage{indentfirst}
\usepackage{dsfont}
\usepackage{amsfonts}
\usepackage{times,amsmath,color,amssymb,epsfig,cite,subfigure,algorithm,algorithmic}
\usepackage{todonotes}
\usepackage{setspace}

\newcommand{\qed}{\nobreak \ifvmode \relax \else
      \ifdim\lastskip<1.5em \hskip-\lastskip
      \hskip1.5em plus0em minus0.5em \fi \nobreak
      \vrule height0.40em width0.6em depth0.25em\fi}

\begin{document}

\title{Covert Waveform Design for Integrated Sensing and Communication System in Clutter Environment}
\author{
\IEEEauthorblockN{ Xuyang Zhao\textsuperscript{1}, Jiangtao Wang\textsuperscript{2}, Xinyu Zhang\textsuperscript{3}, and Yongchao Wang\textsuperscript{4}}
\IEEEauthorblockA{\textsuperscript{1}School of Telecommunication Engineering, Xidian University, China}
\IEEEauthorblockA{\textsuperscript{2}Guangzhou Institute of Technology, Xidian University, China}   \IEEEauthorblockA{\textsuperscript{3}Department of Information and Communications Engineering, Aalto University, Finland}
\IEEEauthorblockA{\textsuperscript{3}State Key Laboratory on ISN, Xidian University, China}
Email: xy\_zhao@stu.xidian.edu.cn}

\maketitle

\begin{abstract}
This paper proposes an integrated sensing and communication (ISAC) system covert waveform design method for complex clutter environments, with the core objective of maximizing the signal-to-clutter-plus-noise ratio (SCNR). The design achieves efficient clutter suppression while meeting the covertness requirement through joint optimization of the transmit waveform and receive filter, enabling cooperative radar detection and wireless communication. This study presents key innovations that explicitly address target Doppler shift uncertainty, significantly enhancing system robustness against Doppler effects. To ensure communication reliability, the method incorporates phase difference constraints between communication signal elements in the waveform design, along with energy constraint, covert constraint, and peak-to-average power ratio (PAPR) constraint. The original non-convex optimization problem is transformed into a tractable convex optimization form through convex optimization technique. Simulation results demonstrate that the optimized waveform not only satisfies the covertness requirement in complex clutter environment, but also achieves superior target detection performance. It also ensures reliable communication and confirms the effectiveness of propose method.
\end{abstract}

\section{Introduction}\label{sec:intro}

With the rapid development of modern wireless communication and intelligent sensing technologies, integrated sensing and communication (ISAC) has become an increasingly important research direction for future wireless networks and intelligent systems \cite{Zhou_22}. Unlike traditional communication and sensing technologies that were typically designed and deployed separately, ISAC aims to achieve deep integration of communication and sensing functions through shared spectrum, hardware, and algorithmic resources. Consequently, ISAC significantly enhances system resource utilization and overall performance\cite{Liu_20a}--\cite{Liu_22a}.

From a feature prioritization perspective, the signal processing in ISAC can be categorized into three approaches: radar-centric design \text{\cite{Gaglione_18}--\cite{Ma_21}}, where communication information is embedded into a radar waveform through time, space, or frequency modulation; communication-centric design \cite{Keskin_21}--\cite{Wei_24}, which involves directly using communication signals such as orthogonal frequency-division multiplexing (OFDM) or orthogonal time-frequency-space (OTFS) for sensing; and joint design, which seeks to provide flexible performance trade-offs between radar and communication. However, due to the inherent open broadcast characteristics of wireless communications, ISAC signals are also vulnerable to detection or eavesdropping, posing security risks.

Covert communication, also known as low probability of detection (LPD) communication, refers to the technology of concealing communication activities. This approach provides a higher level of security and privacy compared to traditional encryption technologies\cite{covert_survey}. Bash et al. conducted a comprehensive analysis of wireless covert transmission performance in additive white Gaussian Noise (AWGN) channel, and established that a transmitter can reliably and covertly transmit $\mathcal{O}(\sqrt{n})$ bits of information to a receiver over $n$ channel uses, which represents the fundamental theoretical limit of covert communication \cite{square-root-law}.

Covert communication has been widely applied across a range of domains. However, its application in the field of ISAC is still at an early stage. In \cite{covert_beam}, the authors maximized radar sensing mutual information (MI) or communication rate under covert constraint to jointly design both sensing and communication beamforming vectors, while ensuring that the base station (BS) can covertly transmit information to legitimate receiver. Addressing ISAC system secrecy rate and radar distance with RIS, the authors in \cite{ISAC_RIS} jointly designed transmission power, bandwidth, and RIS modules with game theory. The authors in \cite{ISAC_Hu} considered a shared beamforming vector for both sensing and communication, taking into account the presence of a multi-antenna warden. The vector was designed by maximizing the communication rate under covert and detection constraints.

The aforementioned studies on the application of covert communication in ISAC assume that Alice transmits Gaussian-codebook signals. However, in practical communication scenarios, the signals transmitted by Alice are typically drawn from discrete constellations, such as phase-shift keying (PSK) modulation. Therefore, how to design waveforms that can simultaneously achieve both sensing requirements and desired communication performance remains an open problem. Furthermore, targets typically move at unknown velocities in practical applications. The resulting Doppler effect introduces phase shifts in the target echoes, leading to matched filter mismatch that severely degrades radar detection performance. Consequently, the designed waveforms must additionally possess Doppler robustness.

In this work, we propose a joint design scheme for transmit waveform and receive filter to address the Doppler frequency offset uncertainty in moving targets. By incorporating a phase difference constraint to ensure reliable transmission of communication information, we formulate a waveform optimization problem aimed at maximizing the radar SCNR. The optimization framework incorporates multiple constraints, including energy constraints, PAPR constraint, waveform amplitude lower-bound constraint, covertness constraint, and phase difference constraint. To address the intractable covertness constraint, an upper bound on the Kullback-Leibler (KL) divergence is derived, and convex optimization techniques are employed to relax the original non-convex optimization problem into a tractable convex formulation for solution. Simulation results demonstrate that the designed waveform not only enhances detection performance against target Doppler uncertainty and satisfies covertness requirements, but also maintains reliable communication performance, thereby validating the superiority of the proposed method.

\section{Problem Formulation}\label{sec:format}
\begin{figure}[htbp]
  \centering
  \centerline{\includegraphics[scale=0.65]{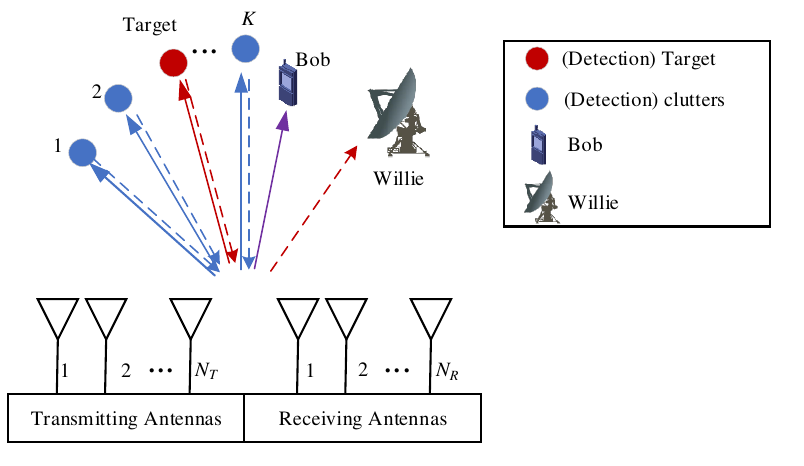}}
\caption{Covert Communication in MIMO ISAC System}
\label{scenario}
\end{figure}
As shown in Fig.~\ref{scenario}, we consider covert communication in a MIMO ISAC system, where the base station (BS) aims to simultaneously detect a target and send information to Bob under the surveillance of a warden, Willie. Assume that the BS is equipped with $N_T$ transmitting antennas and $N_R$ receiving antennas in a uniform linear array (ULA). It transmits an integrated waveform to simultaneously perform target detection and signal transmission to Bob. Willie is assumed to be equipped with a single antenna.

\subsection{Radar Sensing Performance Metric}
The integrated waveform transmitted by the $n_t$-th antenna at time slot $n$ is denoted as ${s_{n_t}}(n)$, where $n_t = 1, \ldots, N_T$, and $n = 1, \ldots, N$. Then the waveform vector transmitted by the $N_T$ antennas at the $n$-th time slot is given by ${\mathbf{s}}(n) = \left[ {s_1(n), s_2(n), \ldots, s_{N_T}(n)} \right]^T \in \mathbb{C}^{N_T \times 1}$. When detecting a target, the echo will be disturbed by $K$ clutters (echos from other angles $\theta_k$). For a target located at azimuth angle $\theta_0$, the echo signal at the $n$-th time slot from the $N_R$ receiving antennas can be expressed as
\begin{equation}\label{xn}
\begin{split}
&{\mathbf{x}}\left( n \right) = {\alpha _0}{{\mathbf{a}}_r}\left( {{\theta _0}} \right){{\mathbf{a}}_t^T}{\left( {{\theta _0}} \right)}{\mathbf{s}}\left( n \right){e^{j2\pi {n} {f_0}}}\\
& \!+\! \sum\limits_{k = 1}^K {{\alpha _k}} {{\mathbf{a}}_r}\left( {{\theta _k}} \right){{\mathbf{a}}_t^T}{\left( {{\theta _k}} \right)}{\mathbf{s}}\left( {n - {r_k}} \right){e^{j2\pi n{f_k(f_0)}}} + {\mathbf{v}}\left( n \right),
\end{split}
\end{equation}
where ${\alpha _0}$ denotes the amplitude of the target echo with $\mathbb{E} [|{\alpha _0}{|^2}] = \sigma _0^2$, while $\alpha_k$ represents the amplitude of the $k$-th clutter component whose second-order moment satisfies $\mathbb{E} [|{\alpha _k}{|^2}] = \sigma _k^2$, where $\alpha_k$ is mutually independent for different $k$. ${f_0}$ represents the normalized Doppler frequency offset of the target, while ${f_k}(f_0)$ represents the normalized Doppler frequency offset of the $k$-th clutter with respect to $f_0$. For the Doppler offset $f_0$ of a moving target, it is difficult to obtain its exact value, and it is assumed that $f_0 \in \Omega$. ${r_k} \in \left\{ {0,1,...,N} \right\}$ represents the delay of the $k$-th clutter. ${\mathbf{v}}\left( n \right)\sim {\mathcal{CN}}({\mathbf{0}},\sigma _v^2{\mathbf{I}})$ denotes the AWGN. ${{\mathbf{a}}_t}\left( \theta \right)$ and ${{\mathbf{a}}_r}\left( \theta \right)$ denote the transmit and receive steering vectors, respectively, which can be expressed as
\begin{equation}\label{a_t}
{{\mathbf{a}}_t}\left( \theta  \right) = \frac{1}{{\sqrt {{N_t}} }}{\left[ {1,{e^{ - j\pi \sin \theta }}, \ldots ,{e^{ - j\pi \left( {{N_t} - 1} \right)\sin \theta }}} \right]^T},
\end{equation}
\begin{equation}\label{a_r}
{{\mathbf{a}}_r}\left( \theta  \right) = \frac{1}{{\sqrt {{N_r}} }}{\left[ {1,{e^{ - j\pi \sin \theta }}, \ldots ,{e^{ - j\pi \left( {{N_r} - 1} \right)\sin \theta }}} \right]^T}.
\end{equation}

To facilitate subsequent matrix operations, we define ${\mathbf{x}} = {[{\mathbf{x}}{(1)^T} \ldots {\mathbf{x}}{(N)^T}]^T}$, ${\mathbf{s}} = {[{\mathbf{s}}{(1)^T} \ldots {\mathbf{s}}{(N)^T}]^T}$, and ${\mathbf{v}} = {[{\mathbf{v}}{(1)^T} \ldots {\mathbf{v}}{(N)^T}]^T}$. Then, $\mathbf{x}$ can be written as
\begin{equation}\label{equationx}
{\mathbf{x}} = {\alpha _0}{\mathbf{A}}\left( {{r_0},{\theta _0},{f_0}} \right){\mathbf{s}} + \sum\limits_{k = 1}^K {{\alpha _k}} {\mathbf{A}}\left( {{r_k},{\theta _k},{f_k}(f_0)} \right){\mathbf{s}} + {\mathbf{v}}.
\end{equation}
The expression of ${\mathbf{A}}\left( {{r_k},{\theta _k},{f_k(f_0)}} \right)$ is given by
\begin{equation}\label{artf}
{\mathbf{A}}\left( {{r_k},{\theta _k},{f_k}(f_0)} \right) \!\!=\!\! \left( {\text{Diag} \left( {{\mathbf{p}}\left( {{f_k}(f_0)} \right)} \right) \!\otimes\! \left( {{{\mathbf{a}}_r}\left( {{\theta _k}} \right){{\mathbf{a}}_t^T}{{\left( {{\theta _k}} \right)}}} \right)} \right){{\mathbf{J}}_{{r_k}}},
\end{equation}
where $ \text{Diag}(\cdot)$ and $ \otimes$ are diagonal matrix and Kronecker product
operators, and ${{\mathbf{J}}_{{r_k}}}(i,j)$ is a shift matrix, denote as
\begin{equation}\label{Jrk}
\begin{aligned}
{{\mathbf{J}}_{{r_k}}}(i,j) = \left\{
\begin{array}{ll}
1, &  i - j = {N_T} \times {r_k}, \\
0, &  i - j \ne {N_T} \times {r_k},
\end{array}
\right.
\end{aligned}
\end{equation}
where $i, j\in \left\{ 1, \ldots ,{N_T}N \right\}$. The vector ${\mathbf{p}}\left( {{f_k}(f_0)} \right)$ is defined as
 \begin{equation}\label{pfk}
{\mathbf{p}}\left( {{f_k}(f_0)} \right) = {\left[ {1,{e^{j2\pi {f_k}(f_0)}}, \ldots ,{e^{j2\pi \left( {N - 1} \right){f_k}(f_0)}}} \right]^T}.
\end{equation}

To simplify the expression, we use ${{\mathbf{A}}_0}(f_0)$ and ${{\mathbf{A}}_k}(f_0)$ to represent ${\mathbf{A}}\left( {{r_0},{\theta _0},{f_0}} \right)$ and ${\mathbf{A}}\left( {{r_k},{\theta _k},{f_k}(f_0)} \right)$.
The signal received by receiver after passing through the receive filter ${{\mathbf{w}_{f_0}}}$ is
\begin{equation}\label{equationr}
s_r = {{\mathbf{w}}_{f_0}^H}{\mathbf{x}} \!= \!{\alpha _0}{{\mathbf{w}}_{f_0}^H}{{\mathbf{A}}_0}(f_0){\mathbf{s}} + {{\mathbf{w}}_{f_0}^H}\sum\limits_{k = 1}^K {{\alpha _k}} {{\mathbf{A}}_k}{\mathbf{s}} \!+\! {{\mathbf{w}}_{f_0}^H}{\mathbf{v}}.
\end{equation}

The output signal-to-clutter plus noise ratio, which is expressed as
\begin{equation}\label{SCNRfswf}
{\operatorname{SCNR}}({\mathbf{s}},{{\mathbf{w}}_{f_0},f_0})
= \frac{{{{\left| {{{\mathbf{w}}_{f_0}^H}{{\mathbf{A}}_0}(f_0){\mathbf{s}}} \right|}^2}{\sigma _0^2}}}{{{\sigma _v^2{\mathbf{w}}_{f_0}^H}({\mathbf{\Psi }}\left( {\mathbf{s},f_0} \right) + {\mathbf{I}}){{\mathbf{w}}}_{f_0}}},
\end{equation}
where
\begin{equation}\label{Psis}
{\mathbf{\Psi }}\left( {\mathbf{s}},f_0\right) = \sum_{k = 1}^K q_k{{\mathbf{A}}_k(f_0)}{\mathbf{s}}{{\mathbf{s}}^H}{[\mathbf{A}_k(f_0)]}^H,
\end{equation}
and ${q_k} = \sigma _k^2/\sigma _v^2$.

\subsection{Communication Performance Metric}
Since the integrated waveform is required to carry information while performing target detection, the signal transmitted to Bob, denoted as $d_{n_t}(n)$, is assumed to be drawn from a finite discrete modulation constellation (e.g, M-PSK) with equal probability. The discrete communication data is represented as ${\mathbf{d}}(n) = \left[ {d_1(n), d_2(n), \ldots, d_{N_T}(n)} \right]^T \in \mathbb{C}^{N_T \times 1}$. To ensure that the integrated waveform can effectively transmit information, we constrain the distortion between the integrated waveform signal $s_{n_t}(n)$ and the communication symbol $d_{n_t}(n)$, so that the original bit information can be correctly demodulated by Bob. Each element of the designed waveform should satisfy the following phase difference constraint
\begin{equation}\label{phase}
\left| {\arg ({s_{n_t}}(n))\! -\! \arg ({d_{n_t}}(n))} \right|\! \leq\! \xi , n \!=\! 1, \cdots ,N; n_t \!=\! 1,\cdots, N_T.
\end{equation}
where $\arg (\cdot)$ denotes the phase of the complex value, and $\xi$ is the preset threshold.

\subsection{Covert Performance Metric}
In the considered scenario, Willie aims to determine whether BS is communication with Bob. Mathematically, Willie needs to perform a hypothesis test between the following two hypotheses
\begin{equation}\label{Willie signals moment}
y_w(n)=\left\{\begin{array}{cc}
 \mathbf{h}^T\mathbf{s}(n)+n_w(n)& \mathcal{H}_1 \\
n_w(n) & \mathcal{H}_0
\end{array}\right.,
\end{equation}
where $n_w(n)\sim\mathcal{CN}(0,\sigma_w^2)$, $\mathbf{h}\in \mathbb{C}^{N_T\times 1}$ is the CSI between the BS and Willie. We assume it to be quasi-static Rayleigh fading channel in this work. Additionally, $\mathcal{H}_0$ represents the null hypothesis that the BS does not send signals, $\mathcal{H}_1$ represents the other hypothesis that the BS sends signals.

In each waveform of $N$ time slots, we have
\begin{equation}\label{Willie signals}
\mathbf{y}_w=\left\{\begin{array}{cc}
 \mathbf{H}\mathbf{s}+\mathbf{n}_w& \mathcal{H}_1 \\
\mathbf{n}_w & \mathcal{H}_0
\end{array}\right.,
\end{equation}
where $\mathbf{n}_w \sim \mathcal{CN}(0,\sigma_w^2\mathbf{I})$, and
\begin{equation}
\mathbf{H}
=\left[\begin{array}{cccc}
\mathbf{h}^T & \mathbf{0}^T & \cdots & \mathbf{0}^T \\
\mathbf{0}^T & \mathbf{h}^T & \cdots & \mathbf{0}^T \\
\vdots & \vdots & \ddots & \vdots \\
\mathbf{0}^T & \mathbf{0}^T & \cdots & \mathbf{h}^T
\end{array}\right].
\end{equation}

According to \eqref{phase}, we can obtain that the probability of each realization of the random variable $\mathbf{s}$ is $p =\frac{1}{M^{N_TN}}$. Then, $f\left(\mathbf{y}_w|\mathcal{H}_1\right)$ can be expressed as
\begin{equation}
\begin{aligned}
    &f\left(\mathbf{y}_w|\mathcal{H}_1\right) = \\
    &\frac{p}{\pi^N\operatorname{det}\left(\sigma_w^2\mathbf{I}\right)}\sum_{r=1}^{M^{N_TN}}\operatorname{exp}\left[-\left(\mathbf{y}_w-\boldsymbol{u}_r\right)^H\left(\sigma_w^2\mathbf{I}\right)^{-1}\left(\mathbf{y}_w-\boldsymbol{\mu}_r\right)\right],
    \end{aligned}
\end{equation}
and $f\left(\mathbf{y}_w|\mathcal{H}_0\right)$ can be expressed as
\begin{equation}
        f\left(\mathbf{y}_w|\mathcal{H}_0\right) = \frac{1}{\pi^N\operatorname{det}\left(\sigma_w^2\mathbf{I}\right)}\operatorname{exp}\left[-\mathbf{y}_w^H\left(\sigma_w^2\mathbf{I}\right)^{-1}\mathbf{y}_w\right],
\end{equation}
where $\boldsymbol{\mu}_r = \mathbf{H}\mathbf{s}_r$, $\mathbf{s}_r$ denotes the integrated waveform optimized for a random realization $\mathbf{d}_r$ of the desired signal of Bob.

According to \cite{square-root-law}, the lower bound of Willie's detection error probability $\xi$ satisfies the following constraint to guarantee covert performance.
\begin{equation}\label{covert performance}
    \xi \geq 1-\sqrt{\frac{1}{2}\mathcal{D}\left(\mathbb{P}_1||\mathbb{P}_0\right)} \geq 1-\epsilon,
\end{equation}
where $\mathcal{D}\left(\mathbb{P}_1||\mathbb{P}_0\right)$ is the KL divergence.

From \eqref{covert performance}, the covert constraint can be formulated as follow
\begin{equation}
    \mathcal{D}\left(\mathbb{P}_1||\mathbb{P}_0\right)\leq2\epsilon^2.
\end{equation}

Specifically,
\begin{equation} \label{KL original}
     \mathcal{D}\left(\mathbb{P}_1||\mathbb{P}_0\right) = \int f\left(\mathbf{y}_w|\mathcal{H}_1\right) \log\frac{f\left(\mathbf{y}_w|\mathcal{H}_1\right)}{f\left(\mathbf{y}_w|\mathcal{H}_0\right)}d\mathbf{y}_w.
\end{equation}

Since the integral \eqref{KL original} is difficult to compute directly, we apply the log-sum inequality to obtain
\begin{equation}
\begin{aligned}
    &\mathcal{D}\left(\mathbb{P}_1||\mathbb{P}_0\right)\\
    &\leq\frac{1}{M^{N_TN}}\sum_{r=1}^{M^{N_TN}} \int f\left(\mathbf{y}_w|\mathcal{H}_1,\boldsymbol{\mu}_r\right) \log\frac{f\left(\mathbf{y}_w|\mathcal{H}_1,\boldsymbol{\mu}_r\right)}{f\left(\mathbf{y}_w|\mathcal{H}_0\right)}d\mathbf{y}_w\\
    &=\frac{1}{M^{N_TN}}\sum_{r=1}^{M^{N_TN}}\boldsymbol{\mu}_r^H\left(\sigma_w^2\mathbf{I}\right)^{-1}\boldsymbol{\mu}_r.
    \end{aligned}
\end{equation}

Then, the covert constraint \eqref{KL original} can be simplified as for each $\boldsymbol{\mu}_r$
\begin{equation}\label{covert constraint} \boldsymbol{\mu}_r^H\left(\sigma_w^2\mathbf{I}\right)^{-1}\boldsymbol{\mu}_r\leq2\epsilon^2.
\end{equation}

The corresponding constraint for integrated waveform design is given by
\begin{equation}
\left(\mathbf{Hs}\right)^H\left(\sigma_w^2\mathbf{I}\right)^{-1}\mathbf{Hs}\leq2\epsilon^2.
\end{equation}

Here, we assume that Willie is aware of $\left(\mathbf{Hs}\right)^H\left(\sigma_w^2\mathbf{I}\right)^{-1}\mathbf{Hs}$, which represents the worst-case scenario from the perspective of covert strategy design.

\subsection{Optimization Problem Formulation}
During the transmission and reception of integrated waveforms for sensing and communication, it is necessary to consider energy constraints and peak-to-average power ratio (PAPR) constraints to ensure the practicality of the waveforms. Furthermore, since higher symbol energy reduces the impact of noise during demodulation, a lower bound constraint on the amplitude of each symbol in the waveform is required. Therefore, based on the above analysis of radar and covert communication requirements, the design problem of integrated covert waveforms for ISAC systems can be formulated as follows
\begin{equation}\label{mode1}
\begin{split}
\mathop {\max }\limits_{{\mathbf{s}},{\mathbf{w}}_{f_0}}
& \min\limits_{f_0 \in \Omega } \dfrac{{|{\mathbf{w}}_{f_0}^H{\mathbf{A}}_0(f_0){\mathbf{s}}|^2}}{{{\mathbf{w}}_{f_0}^H({\mathbf{\Psi }}({\mathbf{s},f_0}) + {\mathbf{I}}){\mathbf{w}}_{f_0}}} \\
s.t.\ & C_1:  \|{\mathbf{s}}_{n_t}\|^2 = N, \\
& C_2:\!  \dfrac{\max\limits_{1 \le n \le N} \{|{s_{n_t}}(n)|^2\}}{\frac{1}{N} \|{\mathbf{s}}_{n_t}\|^2} \le \gamma_{up},\\
& C_3:\!  |{s_{n_t}}(n)| \ge \gamma_{low}, \\
& C_4: \!|\arg({s_{n_t}}(n))\! - \!\arg({d_{n_t}}(n))| \!\le\! \xi,\\
&C_5:\left(\mathbf{Hs}\right)^H\left(\sigma_w^2\mathbf{I}\right)^{-1}\mathbf{Hs}\leq2\epsilon^2,
\end{split}
\end{equation}
where $n = 1, \cdots, N; n_t = 1, \cdots, N_T$, and $C_1-C_5$ are the energy constraint, PAPR constraint, amplitude lower bound constraint, phase difference constraint, and covert constraint, respectively.

By incorporating constraint $C_1$ into PAPR constraint, $C_2$ can be represented as
\begin{equation}
\left| {{s_{n_t}}\left( n \right)} \right|^2  \le {\gamma _{up}}.
\end{equation}

We define $\mathbf{u}_{n_t} = [\overbrace {0, \ldots ,0}^{n_t - 1},1,\overbrace {0, \ldots ,0}^{{N_T} - n_t}]$, ${{\mathbf{s}}_{n_t}} = \left( {{{\mathbf{I}}_N} \otimes {{\mathbf{u}}_{n_t}}} \right){\mathbf{s}} = {{\mathbf{U}}_{n_t}}{\mathbf{s}}$. Then, $C_1$ can be rewritten as
\begin{equation}\label{sm2}
{\left\| {{{\mathbf{s}}_{n_t}}} \right\|^2} = {{\mathbf{s}}^H}{\mathbf{U}}_{n_t}^H{{\mathbf{U}}_{n_t}}{\mathbf{s}} = {{\mathbf{s}}^H}{{\mathbf{R}}_{n_t}}{\mathbf{s}}=N.
\end{equation}

Thus, the optimization problem \eqref{mode1} can be reformulated as follows
\begin{equation}\label{modle2}
\begin{split}
\mathop {\max }\limits_{{\mathbf{s}},{\mathbf{w}}_{f_0}} &  \min\limits_{f_0 \in \Omega } \dfrac{{|{\mathbf{w}}_{f_0}^H{\mathbf{A}}_0(f_0){\mathbf{s}}|^2}}{{{\mathbf{w}}_{f_0}^H({\mathbf{\Psi }}({\mathbf{s},f_0}) + {\mathbf{I}}){\mathbf{w}}_{f_0}}} \\
s.t.\ & {{C_1}:\!{{\mathbf{s}}^H}{{\mathbf{R}}_{n_t}}{\mathbf{s}} = N,} \\
& {{C_2}:\!{{\left| {{s_{n_t}}(n)} \right|}^2} \le {\gamma_{up}}},\\
&C_3:\!  |{s_{n_t}}(n)| \ge \gamma_{low},\\
& C_4: \!\left|\arg({s_{n_t}}(n))\! - \!\arg({d_{n_t}}(n))\right| \!\le\! \xi,\\
&C_5:\left(\mathbf{Hs}\right)^H\left(\sigma_w^2\mathbf{I}\right)^{-1}\mathbf{Hs}\leq2\epsilon^2.
\end{split}
\end{equation}

\section{Joint Waveform and Receive Filter Design}
In this section, an alternating algorithm is proposed to solve problem \eqref{modle2}. The approach alternately optimizes ${{\mathbf{w}}_{f_0}}$ and $\mathbf{s}$.
\subsection{Closed-Form Solution for ${{\mathbf{w}}_f}$}
We first fix the variable ${\mathbf{s}}$, and constrain ${\mathbf{w}}_{f_0}^H{{\mathbf{A}}_0}(f_0){\mathbf{s}} = 1$. Then, problem \eqref{modle2} is transformed into an optimization problem for ${{\mathbf{w}}_f}$, and its optimal solution can be easily obtained as follows
\begin{equation}\label{wf}
{{\mathbf{w}}_{f_0}} = \frac{{{{\left[ {{\mathbf{\Psi }}\left( {\mathbf{s},f_0} \right) + {\mathbf{I}}} \right]}^{ - 1}}{{\mathbf{A}}_0}(f_0){\mathbf{s}}}}{{{{\mathbf{s}}^H}{\mathbf{A}}_0^H(f_0){{\left[ {{\mathbf{\Psi }}\left( {\mathbf{s}, f_0} \right) + {\mathbf{I}}} \right]}^{ - 1}}{{\mathbf{A}}_0}(f_0){\mathbf{s}}}}, \quad f_0 \in\Omega.
\end{equation}
\subsection{Solution for ${\mathbf{s}}$}
Substituting the solution from \eqref{wf} into \eqref{modle2}, we obtain
\begin{equation}\label{modles}
\begin{split}
\mathop {\max }\limits_{\mathbf{s}} & \mathop {\min }\limits_{f_0 \in \Omega } {{\mathbf{s}}^H} {\mathbf{Q}}_{f_0}({\mathbf{s}}) {\mathbf{s}} \\
s.t. \  &C_1: {{{\mathbf{s}}^H}{{\mathbf{R}}_{n_t}}{\mathbf{s}} = N}, \\
&C_2: \left| {{s_{n_t}}(n)} \right|^2 \le {\gamma _{up}},\\
&C_3: \left| {{s_{n_t}}(n)} \right| \ge {\gamma _{low}}, \\
&C_4: \left|\arg({s_{n_t}}(n))\! - \!\arg({d_{n_t}}(n))\right|\leq \xi,\\
&C_5: \left(\mathbf{Hs}\right)^H\left(\sigma_w^2\mathbf{I}\right)^{-1}\mathbf{Hs}\leq2\epsilon^2,
\end{split}
\end{equation}
where
\begin{equation}\label{Qfs}
{\mathbf{Q}}_{f_0}({\mathbf{s}}) = {\mathbf{A}}_{_0}^H(f_0){\left[ {{\mathbf{\Psi }}\left( {\mathbf{s},f_0} \right) + {\mathbf{I}}} \right]^{ - 1}}{{\mathbf{A}}_0}(f_0).
\end{equation}

In order to further simplify the minimax problem \eqref{modles}, we introduce the auxiliary variable $v$ and construct the following problem
\begin{equation}\label{modles2}
\begin{split}
\mathop {\max }\limits_{{\mathbf{s}},v} &\ \ v \\
s.t. \ &C_1: {{\mathbf{s}}^H}{{\mathbf{Q}}_{f_0}}({\mathbf{s}}){\mathbf{s}} \ge v, \   f_0 \in \Omega, \\
     &C_2: {{\mathbf{s}}^H}{{\mathbf{R}}_{n_t}}{\mathbf{s}} = N, \\
    &C_3: \left| {{s_{n_t}}(n)} \right| \ge {\gamma _{low}}, \\
     &C_4: \left| {{s_{n_t}}(n)} \right|^2 \le {\gamma _{up}}, \\
     &C_5: \left|\arg({s_{n_t}}(n))\! - \!\arg({d_{n_t}}(n))\right| \leq \xi, \\
     &C_6: \left(\mathbf{Hs}\right)^H\left(\sigma_w^2\mathbf{I}\right)^{-1}\mathbf{Hs}\leq2\epsilon^2.
\end{split}
\end{equation}

Since the constraints $C_1$--$C_3$ and $C_5$ from problem \eqref{modles2} are non-convex, the problem remains difficult to solve directly.
Then, we focus on dealing with these constraints.
${{\mathbf{Q}}_{f_0}}({\mathbf{s}})$ is a positive semidefinite matrix, and any feasible points ${\mathbf{\tilde s}}$ satisfy
\begin{equation}\label{Qfs2}
{({\mathbf{s}} - {\mathbf{\tilde s}})^H}{{\mathbf{Q}}_{f_0}}({\mathbf{s}})({\mathbf{s}} - {\mathbf{\tilde s}}) \ge 0,  \ f_0 \in \Omega.
\end{equation}

The following inequality is obtained
\begin{equation}\label{Qfs3}
{{\mathbf{s}}^H}{{\mathbf{Q}}_{f_0}}({\mathbf{s}}){\mathbf{s}} \ge 2{\rm{Re}}\left({{\mathbf{s}}^H}{{\mathbf{Q}}_{f_0}}({\mathbf{s}}){\mathbf{\tilde s}}\right) - {{\mathbf{\tilde s}}^H}{{\mathbf{Q}}_{f_0}}({\mathbf{s}}){\mathbf{\tilde s}}, \ f_0 \in \Omega.
\end{equation}

Based on the results of the above inequalities, the non-convex constraint $C_1$ can be approximately relaxed as
\begin{equation}\label{Qfs4}
{{\mathbf{\tilde s}}^H}{{\mathbf{Q}}_{f_0}}({\mathbf{s}}){\mathbf{\tilde s}} - 2{\rm{Re}}({{\mathbf{s}}^H}{{\mathbf{Q}}_{f_0}}({\mathbf{s}}){\mathbf{\tilde s}}) \le - v, \ f_0 \in \Omega.
\end{equation}

The constraint $C_2$ in problem \eqref{modles2} can be written equivalently as follows
\begin{equation}\label{sHRs}
N \le {{\mathbf{s}}^H}{{\mathbf{R}}_{n_t}}{\mathbf{s}} \le N.
\end{equation}

It can be seen that the two inequalities in \eqref{sHRs} have only one point of intersection, which makes subsequent solutions very difficult. Based on the successive convex approximation (SCA)  method, we introduce a relaxation variable $\eta$ to construct the following inequality
\begin{equation}\label{sHRs2}
N - \eta  \le {{\mathbf{s}}^H}{{\mathbf{R}}_{n_t}}{\mathbf{s}} \le N + \eta.
\end{equation}

Since ${{\mathbf{R}}_{n_t}}$ is a positive semidefinite matrix, the right-hand constraint is convex, and the non-convex inequality on the left can be approximated as follow
\begin{equation}\label{sHRs3}
{{\mathbf{\tilde s}}^H}{{\mathbf{R}}_{n_t}}{\mathbf{\tilde s}} - 2{\rm{Re}}({{\mathbf{s}}^H}{{\mathbf{R}}_{n_t}}{\mathbf{\tilde s}}) \le \eta  - N.
\end{equation}

The constraint $C_3$ in \eqref{modles2} can also be approximated as follow
\begin{equation}\label{sHRs4}
\gamma _{low}^2 \!+ \!|{\tilde s_{n_t}}(n){|^2}\! -\! {\rm{Re}}\{ 2{\tilde s_{n_t}^*}{(n)}{s_{n_t}}(n)\} \! \le\! 0.
\end{equation}

For the phase constraint $C_5$, we have
\begin{equation}
\begin{aligned}
&\left(d_{n_t}(n)\right)^*s_{n_t}(n)\\
&= |d_{n_t}(n)||s_{n_t}(n)|\cos\left(\arg({s_{n_t}}(n))\! - \!\arg({d_{n_t}}(n))\right).
\end{aligned}
\end{equation}

Based on the properties of $\cos(\cdot)$ and $\arg(\cdot)$, we have
\begin{equation}\label{geometry_phase}
\begin{aligned}
&\cos\left(\arg({s_{n_t}}(n)) - \arg({d_{n_t}}(n))\right) \\
&=  \cos\left(|\arg({s_{n_t}}(n)) - \arg({d_{n_t}}(n))|\right)\\
&\geq \cos\xi.
\end{aligned}
\end{equation}

Then, the originally intractable phase constraint can be transformed into the following linear constraint
\begin{equation}
    \left(d_{n_t}(n)\right)^*s_{n_t}(n) \geq \sqrt{\gamma_{low}}\cos{\xi}.
\end{equation}

The problem \eqref{modles2} can be relaxed as
\begin{equation}\label{modles3}
\begin{split}
\hspace{-0.2cm}\mathop {\min }\limits_{{\mathbf{{s}}},\eta ,v} & - v + \alpha \eta \\
\hspace{-0.2cm}s.t. \  &C_1:{\left( {{{\mathbf{s}}^t}} \right)^H}{{\mathbf{Q}}_{f_0}}\left( {{{\mathbf{s}}}} \right){{\mathbf{s}}^t}\! -\! 2{\rm{Re}}({{{\mathbf{s}}}^H}{{\mathbf{Q}}_{f_0}}\left( {{{\mathbf{s}}}} \right){{\mathbf{s}}^t}) \!\le\!  - v,   f_0 \!\in\! \Omega, \\
     &C_2: {{{\mathbf{s}}}^H}{{\mathbf{R}}_{n_t}}{\mathbf{s}} \le N + \eta, \    \\
     &C_3: {\left( {{{\mathbf{s}}^t}} \right)^H}{{\mathbf{R}}_{n_t}}{{\mathbf{s}}^t} - 2{\rm{Re}}({{{\mathbf{s}}}^H}{{\mathbf{R}}_{n_t}}{{\mathbf{s}}^t}) \!\le \!\eta\!  -\! N,  \\
     &C_4: \left| {{{s}_{n_t}}(n)} \right|^2 \le \gamma,  \\
     &C_5: \gamma _{low}^2 \!+ \!|{ s_{n_t}^{t} }(n){|^2}\! -\! {\rm{Re}}\{ 2{\left({ {s}_{n_t}^{t}}(n)\right)^*}{{s}_{n_t}}(n)\} \! \le\! 0, \\
     &C_6: \left(d_{n_t}(n)\right)^*s_{n_t}(n) \geq \sqrt{\gamma_{low}}\cos{\xi},\\
     &C_7: \left(\mathbf{H{s}}\right)^H\left(\sigma_w^2\mathbf{I}\right)^{-1}\mathbf{H{s}}\leq2\epsilon^2,\\
     &C_8: \eta \ge 0,
\end{split}
\end{equation}
where $\alpha $ is a weight parameter scaling the penalty term $\eta$, and ${{\mathbf{s}}^t}$ represents the solution obtained at the $t$-th iteration. Problem \eqref{modles3} is a convex optimization problem that can be directly solved using the CVX toolbox in MATLAB.

The proposed alternating optimization algorithm is presented in Algorithm 1.
\begin{algorithm}[h]
\caption{Alternating Optimization Algorithm}
\begin{algorithmic}[1]
\REQUIRE ${{\mathbf{s}}^0}, {\mathbf{d}}, \Omega, {\theta _0}, {\theta _k}, {r_0}, {r_k}, {\sigma _0}, {\sigma _k}, {\sigma _n}, {\gamma _{up}}, {\gamma _{low}}$, \\
\qquad \quad $\varepsilon, \xi, \eta, \varsigma $
\STATE Initialization $t = 1,{{\mathbf{s}}^{{\mathbf{(}}1{\mathbf{)}}}}$.
\REPEAT
\STATE according to \eqref{Psis} equation is obtained ${\mathbf{\Psi }}\left( {{{\mathbf{s}}^{(t)}, f_0}} \right)$ according to \eqref{wf} equation is obtained ${{\mathbf{w}}_{f_0}}^{(t)}$, calculated to get the $t$-th time.
\STATE according to \eqref{Qfs} equation is obtained ${{\mathbf{Q}}_{f_0}}\left( {{{\mathbf{s}}^{(t)}}} \right)$ according to \eqref{modles3} is obtained through the CVX toolbox ${{\mathbf{s}}^{(t + 1)}}$.
\STATE Calculated ${\operatorname{SCNR}^{(t + 1)}}$.
\UNTIL $\left| {\operatorname{SCNR}^{(t + 1)}} - \operatorname{SCNR}^{(t)} \right| \le \varsigma $.
\RETURN ${{\mathbf{w}}_{f_0}^ *} ,{{\mathbf{s}}^ * }$
\end{algorithmic}
\end{algorithm}
\section{Numerical Results}
In this section, numerical simulations are performed to evaluate the algorithm performance.
It is assumed that both the transmitter and receiver are ULAs with half-wavelength spacing. ${N_T}{\rm{ }} = {\rm{ }}8,{\rm{ }}{N_R}{\rm{ }} = {\rm{ }}8,{\rm{ }}N{\rm{ }} = {\rm{ }}32$, the angle of the target of interest is ${\theta _0} = {0^ \circ }$ , the discrete delay is ${r_0} = 0$ , power ${\sigma _0}{^2} = 15\text{dBm}$. The angles of the three fixed interferences are ${\theta _1} =  - \frac{\pi }{3}$, ${\theta _2} =  - \frac{\pi }{3}$ and ${\theta _3} = 0$, the discrete delays are ${r_1} = 0$, ${r_2} = -2$, ${r_3} = 2$, the power of each interference is $\sigma _k^2 = 15\text{dBm}$. $d_{n_t}(n)$ is drawn from QPSK modulation constellation. The noise variance is $\sigma_w^2 = \sigma _v^2= -90\text{dBm}$.
\begin{figure}[h]
\centering
\includegraphics[scale=0.6]{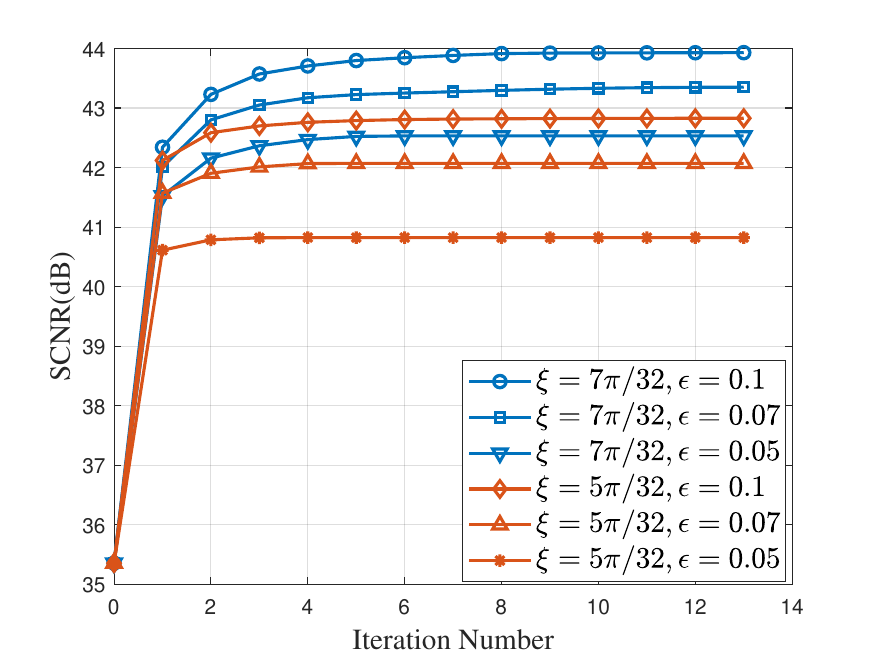}
\caption{Convergence of the objective function with different $\xi$ and $\epsilon$}
\label{SCNR}
\end{figure}

Fig.~\ref{SCNR} presents the convergence performance of the objective function under various phase difference constraints $\xi$ and covertness constraints $\epsilon$. The results demonstrate that the objective function converges successfully, with the SCNR value increasing as both $\xi$ and $\epsilon$ increase. This behavior results from the inherent tradeoff between sensing performance, communication performance and covertness performance. The parameters $\xi$ and $\epsilon$ govern the size of the feasible solution set, where larger values enable the acquisition of waveforms with higher SCNR more readily.
\begin{figure}[h]
\centering
\includegraphics[scale=0.6]{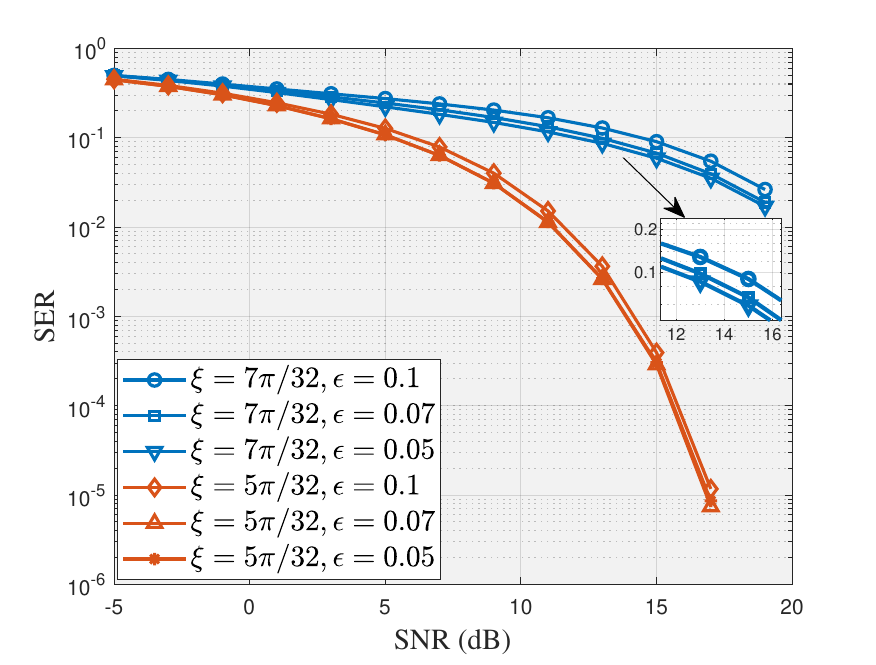}
\caption{Symbol error rate for different $\xi$ and $\epsilon$}
\label{SER}
\end{figure}

Fig.~\ref{SER} presents the symbol error rate (SER) performance under different phase difference constraints $\xi$ and covertness constraints $\epsilon$. It can be observed that the SER decreases with the increase of phase difference $\xi$, which directly affects signal recovery at the receiver. Meanwhile, as the covertness requirement becomes stricter (i.e., $\epsilon$ decreases), the SER deteriorates, demonstrating the inherent trade-off between covertness performance and communication performance.
\begin{figure}[h]
\centering
\includegraphics[scale=0.6]{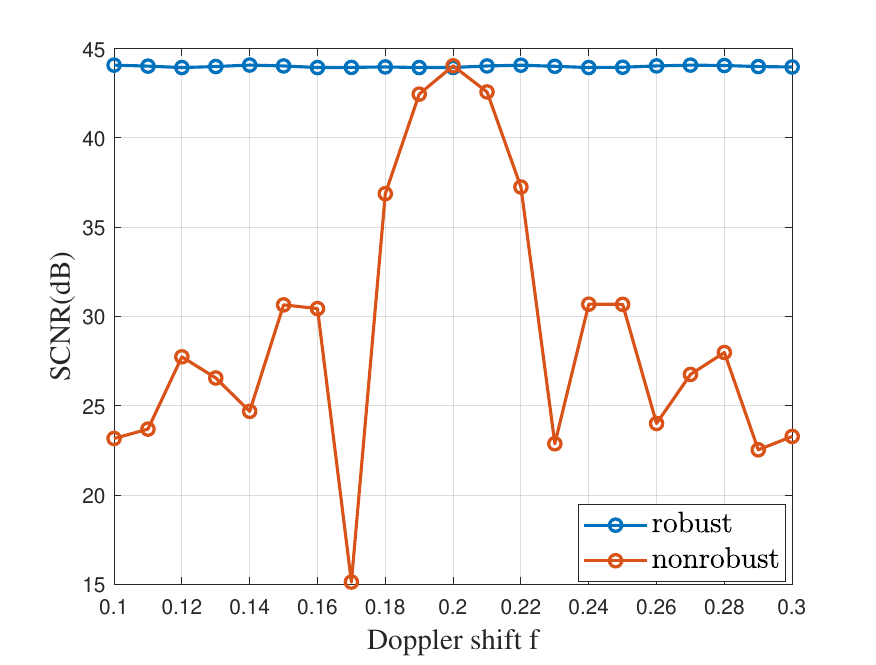}
\caption{Effect of uncertainty of target Doppler offset on SCNR}
\label{Doppler}
\end{figure}

Fig.~\ref{Doppler} demonstrates that the proposed waveform design maintains consistently high SCNR across varying Doppler shifts, confirming its strong robustness over a wide range of target Doppler frequency offsets.

\section{Conclusion}\label{sec:Conclusion}

This paper investigates the covert waveform design for Doppler-resilient ISAC systems in cluttered environments, achieving dual functionality of sensing and communication through phase difference constraints. By formulating the covert waveform design as an SCNR maximization problem with multiple constraints, a customized alternating optimization-based algorithm is proposed to efficiently solve this optimization problem. Simulation results demonstrated that the designed covert waveform maintains excellent detection performance while simultaneously ensuring reliable low bit-error-rate communication and meeting covertness requirements, thereby achieving effective integration of sensing and communication functionalities in covert scenarios.

\bibliographystyle{IEEEbib}

\end{document}